\documentclass[reprint,prc,showpacs,amsmath,amssymb,superscriptaddress]{revtex4-1}
\usepackage{dcolumn}
\usepackage[dvips]{graphicx}
\usepackage{amsmath}
\usepackage{multirow}
\usepackage{setspace}
\newcommand\T{\rule{0pt}{3.1ex}}
\usepackage{color}
\usepackage{bm}
\usepackage[T1]{fontenc}
\usepackage[latin9]{inputenc}
\usepackage{color}
\usepackage{amssymb}
\usepackage{float}

\providecommand{\tabularnewline}{\\}

\usepackage{array}

\begin{document}

\title{Limits on sterile neutrino contributions to neutrinoless double beta decay} 

\author{J.\ Barea}
\email{jbarea@udec.cl}
\affiliation{Departamento de F\'{i}sica, Universidad de Concepci\'{o}n,
 Casilla 160-C, Concepci\'{o}n 4070386, Chile}

\author{J. Kotila}
\email{jenni.kotila@yale.edu}
\affiliation{University of Jyvaskyla, Department of Physics, B.O. Box 35, FI-40014, University of Jyvaskyla, Finland}
\affiliation{Center for Theoretical Physics, Sloane Physics Laboratory,
Yale University, New Haven, Connecticut 06520-8120, USA}
\author{F. Iachello}
\email{francesco.iachello@yale.edu}
\affiliation{Center for Theoretical Physics, Sloane Physics Laboratory,
Yale University, New Haven, Connecticut 06520-8120, USA}

\begin{abstract}
Nuclear matrix elements (NME) for exchange of arbitrary mass neutrinos are calculated in the interacting boson model (IBM-2). By combining the NME with the phase space factors (PSF), expected half-lives for neutrinos of mass $m_N$ and coupling $U_{eN}$ are estimated. Limits on sterile neutrinos with masses in the eV, keV, MeV-GeV, and TeV range are given.

\end{abstract}

\pacs{14.60.Pq, 14.60.St, 23.40.Hc, 23.40.Bw}
\keywords{}
\maketitle

\section{Introduction}
In recent years, the possible occurrence of sterile neutrinos with mass $m_N$ in addition to the three  known species $m_1, m_2, m_3$, has attracted considerable attention, and searches are under way to detect their presence in oscillation experiments and in accelerator experiments. Sterile neutrinos were introduced by Pontecorvo \cite{pon68} in 1968 as neutrinos with no standard model interaction.
 Several suggestions have been made for, among others, sterile neutrinos in the eV mass range \cite{giu13, bar11}, in the keV mass range \cite{asa05}, in the MeV-GeV mass range \cite{sha06, asa11}, and in the TeV mass range \cite{tel11}. 

Sterile neutrinos, if they exist, will contribute to neutrinoless double beta decay (DBD). It is therefore of interest at the present time to estimate the expected half-life for Majorana neutrinos of arbitrary mass. In a previous series of papers we have considered the case of very light, $m_N\ll p_F$, and very heavy $m_N \gg p_F$, neutrinos, where $p_F \sim 100$MeV is the nucleon Fermi momentum in the nucleus (in this article we use units $c=1$). In these cases the half-life factorizes to 
\begin{equation}
\label{hl0}
\lbrack\tau_{1/2}^{0\nu}]^{-1}=G_{0\nu}\left\vert M_{0\nu}\right\vert ^{2}\left\vert f(m_{i},U_{ei})\right\vert ^{2},
\end{equation}
where $G_{0\nu}$ is a phase space factor (PSF), $M_{0\nu}$ the nuclear matrix element (NME), and $f$ is equal to, for light neutrino exchange,
\begin{equation}
f=\frac{\langle m_\nu\rangle}{m_e}, \hspace{1cm}
\langle m_{\nu}\rangle =\sum_{k=light}\left(U_{ek}\right)^{2}m_{k},
\end{equation}
and, for heavy neutrino exchange
\begin{equation}
f  =  m_{p}\left\langle m_{\nu_{h}}^{-1}\right\rangle, \hspace{0.75cm} \langle m_{\nu_{h}}^{-1}\rangle=\sum_{k_h=heavy}\left(U_{ek_h}\right)^{2}\frac{1}{m_{k_h}}.
\end{equation}
When the mass $m_N$ is intermediate, and especially, when it is of the order of magnitude of $p_F$, the factorization (1) is not possible, and physics beyond the standard model is entangled with nuclear physics. In this case, the half-life can be written as
\begin{equation}
\label{hl0inter}
\lbrack\tau_{1/2}^{0\nu}]^{-1}=G_{0\nu}\left\vert \sum_N (U_{eN})^2
 M_{0\nu}(m_N)\frac{m_N}{m_e}\right\vert ^{2}.
\end{equation}

The matrix elements $ M_{0\nu}(m_N)m_N/m_e$ have been calculated within the framework of the interacting shell model, ISM, \cite{ble10} and quasiparticle random phase approximation, QRPA, \cite{fae14}. Here we present results within the microscopic interacting boson model, IBM-2, with isospin restoration \cite{bar15}. From the NME and the PSF that we have previously tabulated \cite{kot12}, we then make estimates for half-lives and set limits on some of the suggested sterile neutrino species.

\section{Formalism}
Although several formulations of $0\nu\beta\beta$ decay are available, we use here the formulation of \v{S}imkovic {\it et al.} \cite{sim99}, as adapted to the case of neutrinos of arbitrary mass, $m_N$ \cite{fae14}. The transition operator $T(p)$ is the same as in Eq.~(5) of our previous paper \cite{bar13} with $f=m_N/m_e$ and 
\begin{equation}
v(p)=\frac{2}{\pi}\frac{1}{\sqrt{p^2+m_{N}^2}\left( \sqrt{p^2+m_{N}^2}+\tilde{A}\right)},
\end{equation}
where $v(p)$ is the so called neutrino "potential". The product $fv(p)$, 
\begin{equation}
fv(p)=\frac{m_{N}}{m_e}\frac{2}{\pi}
\frac{1}{\sqrt{p^2+m_{N}^2}\left( \sqrt{p^2+m_{N}^2}+\tilde{A}\right)}
\end{equation}
has the limits:
\begin{equation}
\begin{split}
m_{N} \rightarrow 0 \hspace{0.25cm} &:\hspace{0.25cm}fv=\frac{m_{N}}{m_e}\frac{2}{\pi}
\frac{1}{p\left(p+\tilde{A}\right)}\\
m_{N} \rightarrow \infty \hspace{0.25cm}&: \hspace{0.25cm}fv=\frac{m_{N}}{m_e}\frac{2}{\pi}
\frac{1}{m_{N}^2}=
\frac{m_p}{m_{N}}
\left(\frac{2}{\pi}
\frac{1}{m_em_{p}}\right)
\end{split}
\end{equation}
as in our previous calculations for light and heavy neutrinos \cite{bar15,bar13}. In Eq.~(5) $\tilde{A}$ is the closure energy assumed to be a smoothly varying function of the mass number $A$ as given in Table 25 of Ref. \cite {tom91}. All other quantities, form factors $\tilde{h}(p)$, momentum dependence of the coupling constants $g_V(p^2)$ and $g_A(p^2)$, are the same as in \cite{bar15,bar13}. Also, in this paper we use Argonne short-range correlation \cite{sim09} as in our previous article \cite{bar15}.

The nuclear matrix elements appearing in Eq.~(4) can be written as
\begin{equation}
\begin{split}
M_{0\nu} (m_N) = &  g_{A}^{2}M^{(0\nu)}(m_N), \\
M^{(0\nu)} (m_N) =&   M_{GT}^{(0\nu)}(m_N)-\left(\frac{g_{V}}{g_{A}}\right)^{2}M_{F}^{(0\nu)}(m_N)
\\&+M_{T}^{(0\nu)}(m_N),
\end{split}
\end{equation}
with the axial vector coupling constant $g_A$ explicitly written in front, where $M^{(0\nu)} (m_N)$ is the dimensionless matrix element calculated with the neutrino potential of Eq.(5).

\section{Results}
\begin{figure*}[ctb!]
\begin{center}
\begin{tabular}{ccc}
\includegraphics[width=0.33\linewidth]{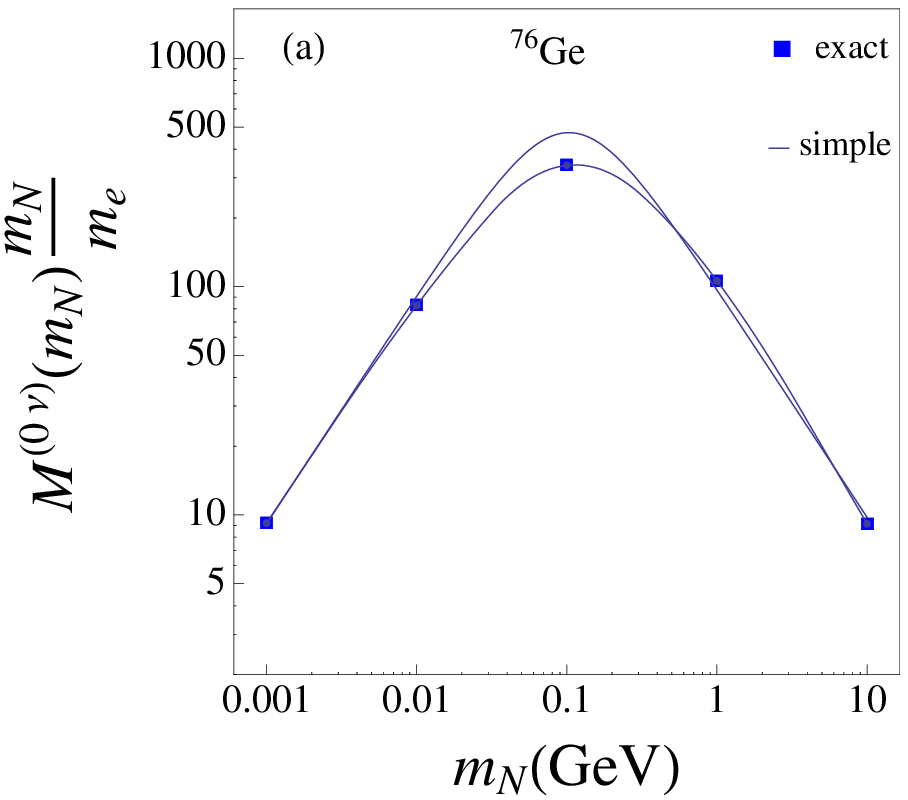} 
&\includegraphics[width=0.33\linewidth]{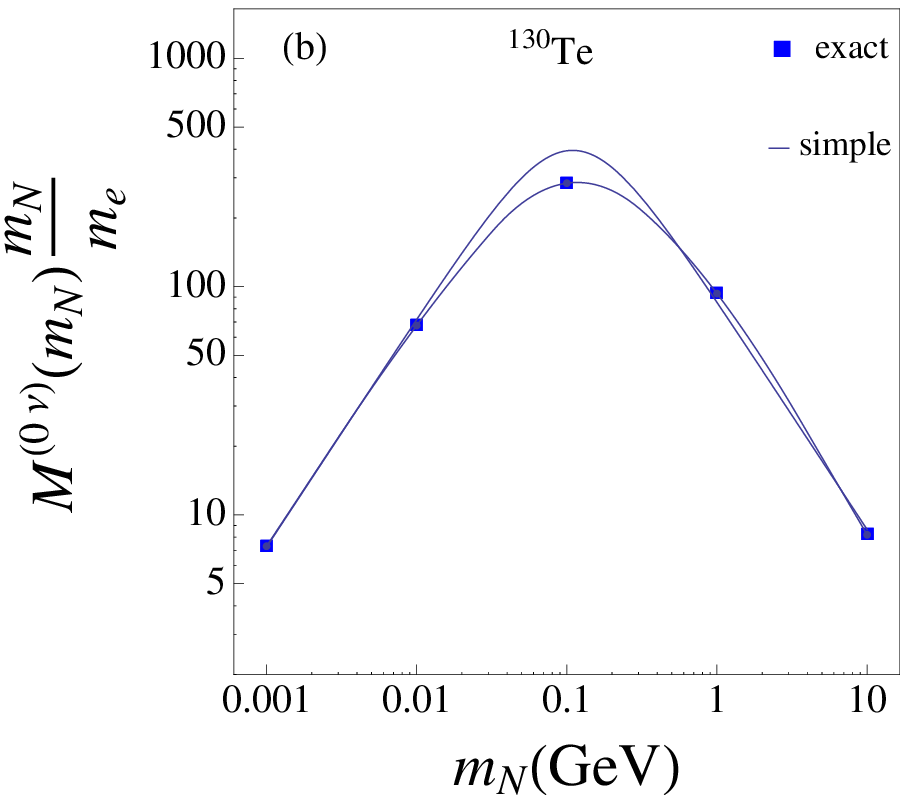} 
&\includegraphics[width=0.33\linewidth]{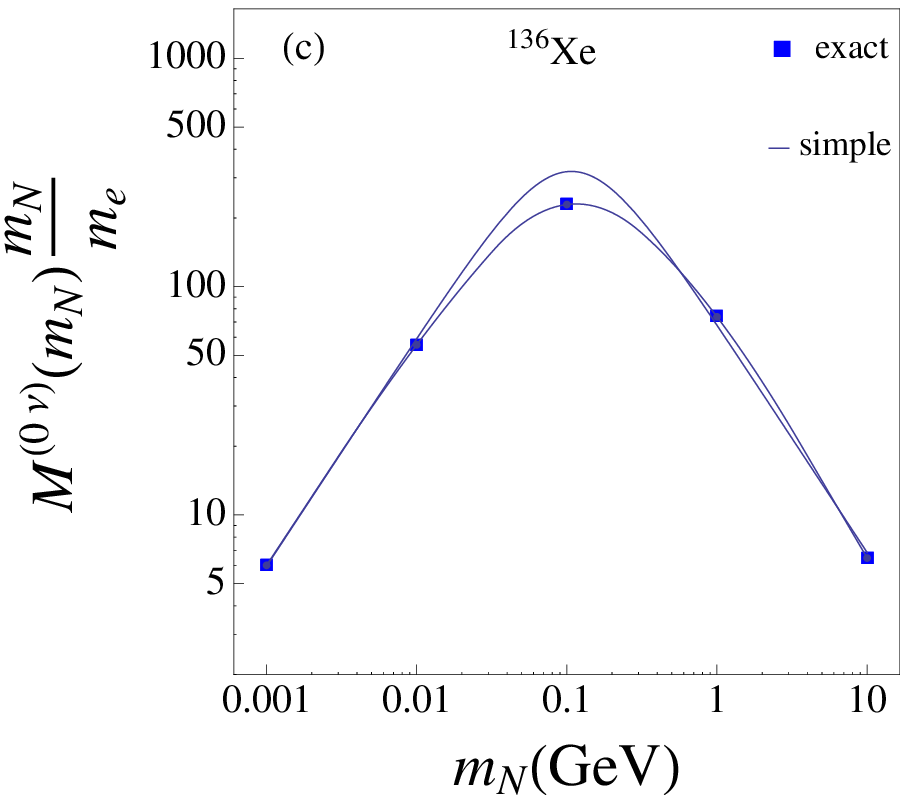} 
\end{tabular}
\caption{\label{fig1}(Color online) IBM-2 NMEs for neutrinos of arbitrary mass plotted as a function of $m_N$ in a) $^{76}$Ge, b) $^{130}$Te, and c) $^{136}$Xe. Blue squares represent the exact calculation for $m_N=0.001$GeV, $0.01$GeV, $0.1$GeV, $1$GeV, $10$GeV, joined together by a Mathematica interpolating formula. The curve is obtained using the simple formula (\ref{simple}).}
\end{center}
\end{figure*}

In Table \ref{table1} we show the dimensionless IBM-2 nuclear matrix elements $M^{(0\nu)} (m_N){m_N/m_e}$ for $0\nu\beta\beta$ decay with Argonne-SRC and $g_V/g_A=1/1.269$ to the ground state, $0^+_1$, and first excited $0^+$ state, $0^+_2$, broken down into Gamow-Teller, GT, Fermi, F, and Tensor, T, contribution and their sum, according to Eq.~(8) for several  values of $m_N$. These matrix elements are plotted in Fig. \ref{fig1} for three isotopes, $^{76}$Ge, $^{130}$Te, and $^{136}$Xe of interest to GERDA, CUORE and KamLAND-Zen/EXO experiments. 

\begin{table*}[cbt!]
 \caption{\label{table1}IBM-2 nuclear matrix elements $M^{(0\nu)}m_N/m_e$ (dimensionless) for $0\nu\beta^-\beta^-$ decay with Argonne SRC, $g_V/g_A=1/1.269$, as a function of $m_N$ in $^{76}$Ge, $^{130}$Te, and $^{136}$Xe.}
 \begin{ruledtabular} %
\begin{tabular}{cccccccccc}
&$m_N$ &\multicolumn{4}{c}{$0_{1}^{+}$} &\multicolumn{4}{c}{$0_{2}^{+}$}\\\cline{3-6}\cline{7-10}
\T
$A$  & (GeV)& $M_{GT}^{(0\nu)}\frac{m_N}{m_e}$  & $M_{F}^{(0\nu)}\frac{m_N}{m_e}$  & $M_{T}^{(0\nu)}\frac{m_N}{m_e}$  & $M^{(0\nu)}\frac{m_N}{m_e}$  & $M_{GT}^{(0\nu)}\frac{m_N}{m_e}$  & $M_{F}^{(0\nu)}\frac{m_N}{m_e}$  & $M_{T}^{(0\nu)}\frac{m_N}{m_e}$  & $M^{(0\nu)}\frac{m_N}{m_e}$\tabularnewline
\hline 
\T
$^{76}$Ge  &0.00001	& 0.0891  & -0.0134  & -0.00494  & 0.0925  		& 0.0386  &-0.00520   & -0.00203  & 0.03980\tabularnewline
$^{76}$Ge  &0.0001		& 0.891  & -0.134  & -0.0494  & 0.92  		& 0.386  &-0.0520   & -0.0203  & 0.398\tabularnewline
$^{76}$Ge  &0.001	& 8.89  & -1.34  & -0.494  & 9.22  		& 3.58  &-0.520   & -0.203  & 3.97\tabularnewline
$^{76}$Ge  &0.010	& 81.4  & -13.3  & -4.92  &82.8   		& 35.1  &-5.16   & -2.01  & 36.3\tabularnewline
$^{76}$Ge  &0.100	& 320  & -91.0  & -36.4 & 340 			& 130  &-34.5   & -14.3  & 137\tabularnewline
$^{76}$Ge  &1.000	& 108  & -39.5  & -26.6  & 106  			& 40.1  &-13.8   & -9.66  & 39.1\tabularnewline
$^{76}$Ge  &10.00	& 9.59  & -4.01  & -2.95  & 9.14  		& 3.53  &-1.40   & -1.07  & 3.34\tabularnewline
$^{76}$Ge  &100.0	& 0.957  & -0.402  & -0.295  & 0.911  		& 0.353  &-0.140   & -0.107  & 0.333\tabularnewline
$^{76}$Ge  &1000	& 0.0957  & -0.0402  & -0.0295  & 0.0911  		& 0.0353  &-0.0140   & -0.0107  & 0.0333\tabularnewline
\hline
\T
$^{130}$Te  &0.00001	& 0.0681  & -0.0127  & -0.00285  & 0.0731  		& 0.0499 & -0.00878  & -0.00174  & 0.0536\tabularnewline
$^{130}$Te  &0.0001		& 0.681  & -0.127  & -0.0285  & 0.73 1 			& 0.499 & -0.0878  & -0.0174  & 0.536\tabularnewline
$^{130}$Te  &0.001		& 6.79  & -1.27  & -0.285  & 7.30  				& 4.97 & -0.878  & -0.174  & 5.34\tabularnewline
$^{130}$Te  &0.010		& 62.6  & -12.6  & -2.83  & 67.6  				& 45.6 & -8.70  & -1.73  & 49.2\tabularnewline
$^{130}$Te  &0.100		& 253  & -82.8  & -21.2  & 284  					& 172 & -54.9  & -12.5  & 194\tabularnewline
$^{130}$Te  &1.000		& 88.0  & -34.9  & -16.4  & 93.3  				& 53.3 & -21.0  & -9.18  & 57.2\tabularnewline
$^{130}$Te  &10.00		& 7.81  & -3.55  & -1.82  & 8.19  				& 4.70 & -2.12  & -1.02  & 5.00\tabularnewline
$^{130}$Te  &100.0		& 0.779  & -0.356  & -0.183  & 0.818  			& 0.469 & -0.212  & -0.102  & 0.499\tabularnewline
$^{130}$Te  &1000		& 0.0779  & -0.0356  & -0.0183  & 0.0818  		& 0.0469 & -0.0212  & -0.0102  & 0.0499\tabularnewline
\hline
\T
$^{136}$Xe  &0.00001	& 0.0562  & -0.0102  & -0.00216  & 0.0603 		& 0.0294 &-0.00468   & -0.000755  & 0.0316\tabularnewline
$^{136}$Xe  &0.0001	& 0.562  & -0.102  & -0.0216  & 0.603  		& 0.294 &-0.0468   & -0.00755  & 0.316\tabularnewline
$^{136}$Xe  &0.001	& 5.60  & -1.02  & -0.216  & 6.02  		& 2.94 &-0.468   & -0.0755  & 3.15\tabularnewline
$^{136}$Xe  &0.010	& 51.5  & -10.1  & -2.16  & 55.6  		& 26.7 &-4.64   & -0.755  & 28.8\tabularnewline
$^{136}$Xe  &0.100	& 204  & -65.8  & -16.1  & 229 			& 93.4 &-28.6   & -5.43  & 106\tabularnewline
$^{136}$Xe  &1.000	& 69.4  & -27.4  & -12.5  & 73.8  		& 26.4 &-10.3   & -4.02  & 28.8\tabularnewline
$^{136}$Xe  &10.00	& 6.15  & -2.79  & -1.40  & 6.48  		& 2.32 &-1.03   & -0.448  & 2.51\tabularnewline
$^{136}$Xe  &100.0	& 0.614  & -0.279  & -0.140  & 0.647  		& 0.232 &-0.103   & -0.0449  & 0.251\tabularnewline
$^{136}$Xe  &1000	& 0.0614  & -0.0279  & -0.0140  & 0.0647  		& 0.0232 &-0.0103   & -0.00449  & 0.0251\tabularnewline

\end{tabular}
\end{ruledtabular} 
\end{table*}

The interesting aspect of Fig. \ref{fig1} is that the curves peak at $m_N\sim100$MeV, the scale set by the nucleon Fermi momentum in the nucleus, $p_F$. If sterile neutrinos of this mass exist, their contribution to the half-life is enhanced. The relatively simple behavior of Fig. \ref{fig1} allows one to write a simple formula to estimate the effect of sterile neutrinos on half-lives for $0\nu\beta\beta$ decay. The simple formula is 
\begin{equation}
\label{simple}
\lbrack\tau_{1/2}^{0\nu}]^{-1}=G_{0\nu}g_A^4\left\vert M^{(0\nu_h)}\right\vert^2
\left\vert m_p\sum_N (U_{eN})^2\frac{m_N}{\langle p^2\rangle+ m_N^2}
 \right\vert ^{2},
\end{equation}
with
\begin{equation}
\langle p^2\rangle=\frac{M^{(0\nu_h)}}{M^{(0\nu)}}m_pm_e,
\end{equation}
where $M^{(0\nu)}$ and $M^{(0\nu_h)}$ are the matrix elements calculated in the limits $m_N \rightarrow 0$ and $m_N \rightarrow \infty$, respectively.
The simple formula (\ref{simple}) is identical to that given in \cite{fae14} and suggested in \cite{kov09}. It provides a good approximation to the exact calculation, except in the region $m_N\sim p_F$, as one can see in Fig. \ref{fig1} where the simple formula (\ref{simple}) is compared with the exact calculation. The values of $M^{(0\nu)}$ and $M^{(0\nu_h)}$ were given in our previous calculation \cite{bar15} and are reproduced in Table \ref{table2}, together with the value of $\langle p^2\rangle$ for all nuclei of interest. The error in this table has been increased from 16\% and 28\%  in \cite{bar15} to 32\% and  56\% for $M^{(0\nu)}$ and $M^{(0\nu_h)}$, respectively, to account for a SRC different from Argonne. SRC greatly affect the value of $M^{(0\nu_h)}$ as discussed in Ref. \cite{bar13}.
 \begin{table}[h]
 \caption{\label{table2}The matrix elements $M^{(0\nu)}$, $M^{(0\nu_h)}$ and the value of $\langle p^2\rangle$ for all nuclei of interest in DBD.}
 \begin{ruledtabular} %
\begin{tabular}{cccc}
Decay  & $M^{(0\nu)}$ & $M^{(0\nu_h)}$  & $\sqrt{\langle p^2\rangle}$ (MeV)\tabularnewline
\hline 
\T
$^{48}$Ca  &1.75(56)		&47(26)		&113  \tabularnewline
$^{76}$Ge  &4.68(150)		&104(58)		&103		   	 \tabularnewline
$^{82}$Se  &3.73(120)		&83(46)		&103		   	 \tabularnewline
$^{96}$Zr	&2.83(90)		        &99(55)		&130		   	 \tabularnewline
$^{100}$Mo &4.22(135)		&164(92)		&137	   	  \tabularnewline
$^{110}$Pd &4.05(130)		&154(86)		&135		   	 \tabularnewline
$^{116}$Cd &3.10(100)		&110(62)		&130		   	 \tabularnewline
$^{124}$Sn &3.19(102)		&79(44)		&109		   	 \tabularnewline
$^{128}$Te &4.10(131)		&101(57)		&109		   	 \tabularnewline
$^{130}$Te &3.70(118)		&92(52)		&109		   	 \tabularnewline
$^{134}$Xe &4.05(130)		&91(51)		&104	   	  \tabularnewline
$^{136}$Xe &3.05(98)		&73(41)		&107	   	  \tabularnewline
$^{148}$Nd &2.31(74)		&103(58)		&146		   	 \tabularnewline
$^{150}$Nd &2.67(85)		&116(65)		&144		   	 \tabularnewline
$^{154}$Sm &2.82(90)		&113(63)		&139		   	 \tabularnewline
$^{160}$Gd &4.08(131)		&155(87)		&135		   	 \tabularnewline
$^{198}$Pt &2.19(70)		&104(58)		&151		   	 \tabularnewline
$^{232}$Th &4.04(128)		&159(89)		&138	   	 \tabularnewline
$^{238}$U &4.81(154)		&189(106)		&137		   	 \tabularnewline
\end{tabular}\end{ruledtabular} 
\end{table}

\begin{figure*}[ctb!]
\begin{center}
\begin{tabular}{ccc}
\includegraphics[width=0.33\linewidth]{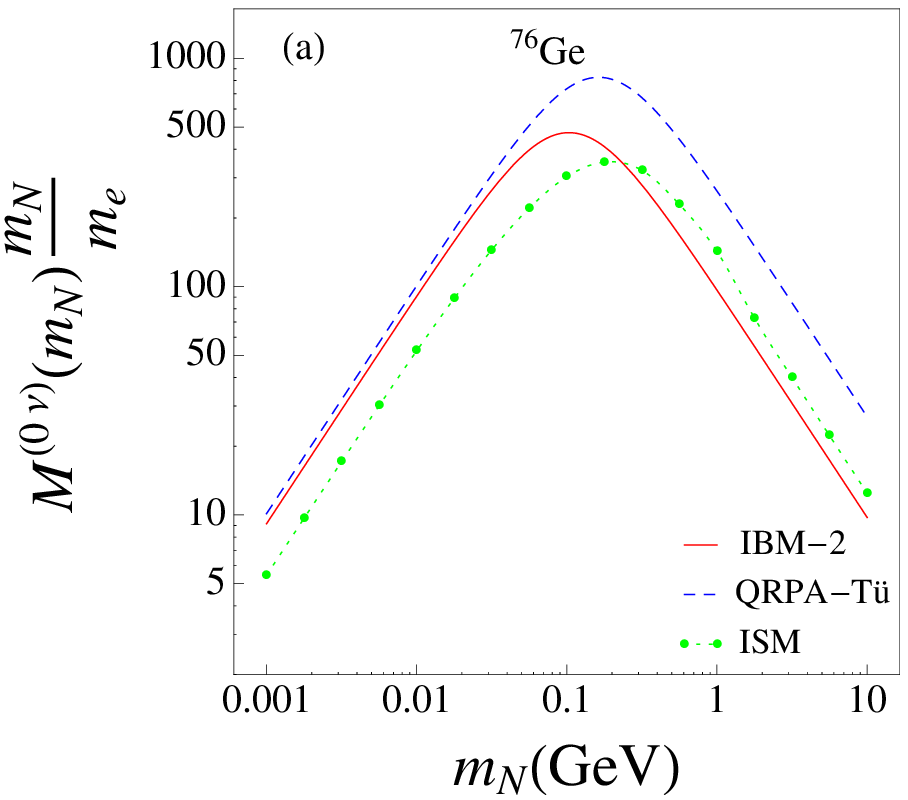} 
&\includegraphics[width=0.33\linewidth]{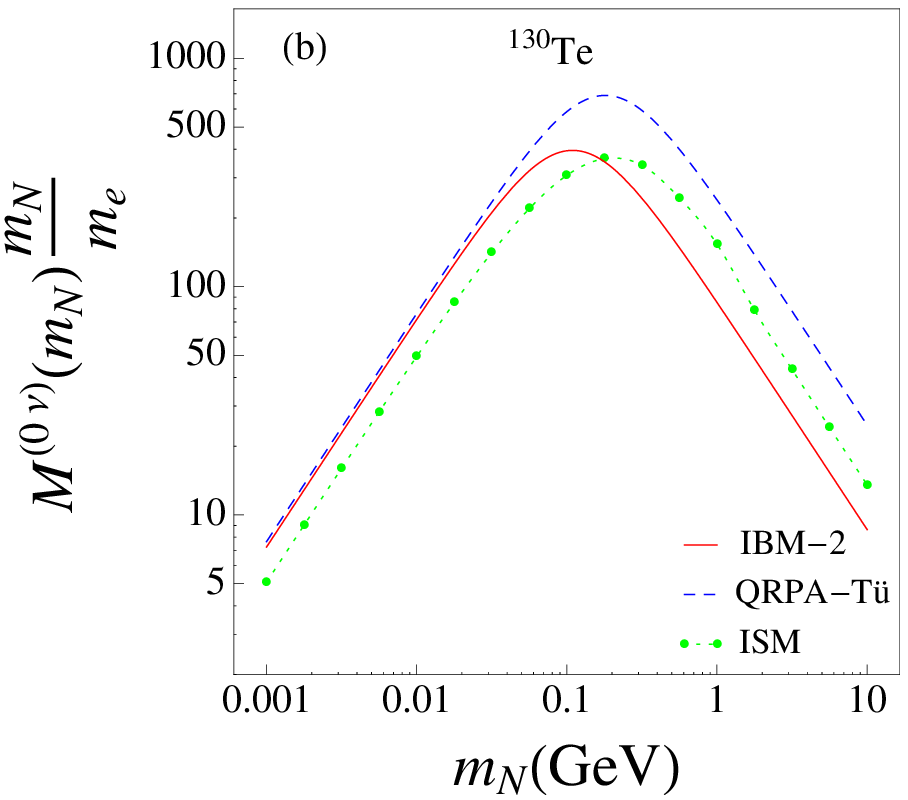} 
&\includegraphics[width=0.33\linewidth]{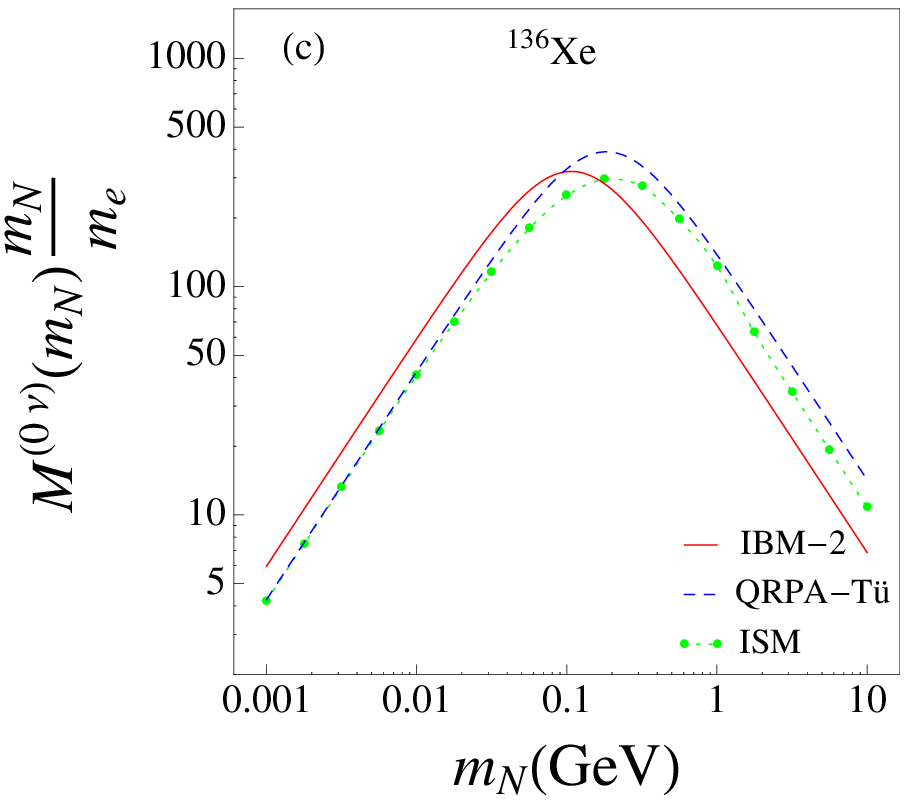} 
\end{tabular}
\caption{\label{newfig2}(Color online) Comparison between IBM-2 (Argonne SRC) (red), QRPA-T{\"u} (Argonne SRC) \cite{fae14} (blue), and ISM (UCOM SRC) (green) NMEs for neutrinos of arbitrary mass plotted as a function of $m_N$ in a) $^{76}$Ge, b) $^{130}$Te, and c) $^{136}$Xe. }
\end{center}
\end{figure*}

It is of interest to compare our results with those of the interacting shell model (ISM) \cite{ble10} and the quasiparticle random phase approximation (QRPA) \cite{fae14}. In Fig. \ref{newfig2} we show this comparison. All three calculations have the same behavior. However, the maximum of the curves, related to the value of the Fermi momentum, $p_F$, is different. For example, in $^{76}$Ge the maximum of the IBM-2 curve is at $m_N\sim 100$MeV, while for QRPA-T{\"u} it is at 160MeV and for ISM at 180MeV. While, for $m_N < p_F$ QRPA and IBM-2 agree, for $m_N>p_F$ the agreement is  between IBM-2 and ISM. The actual form of the curve arises from the interplay between short range correlations, SRC, and the wave functions close to the origin. Also, while the tensor component $M_T$ plays a small role for $m_N<p_F$, it plays a large role for $m_N>p_F$, as one can see from Table I. Because of the divergence of the tensor component at the origin, the numerical evaluation of its matrix elements is very sensitive. For this reason we have increased the IBM-2 error estimate of $M^{(0\nu_h)}$ in Table II.

\section{Half-lives and limits on sterile neutrino contributions}
\begin{figure*}[ctb!]
\begin{center}
\begin{tabular}{ccc}
\includegraphics[width=0.33\linewidth]{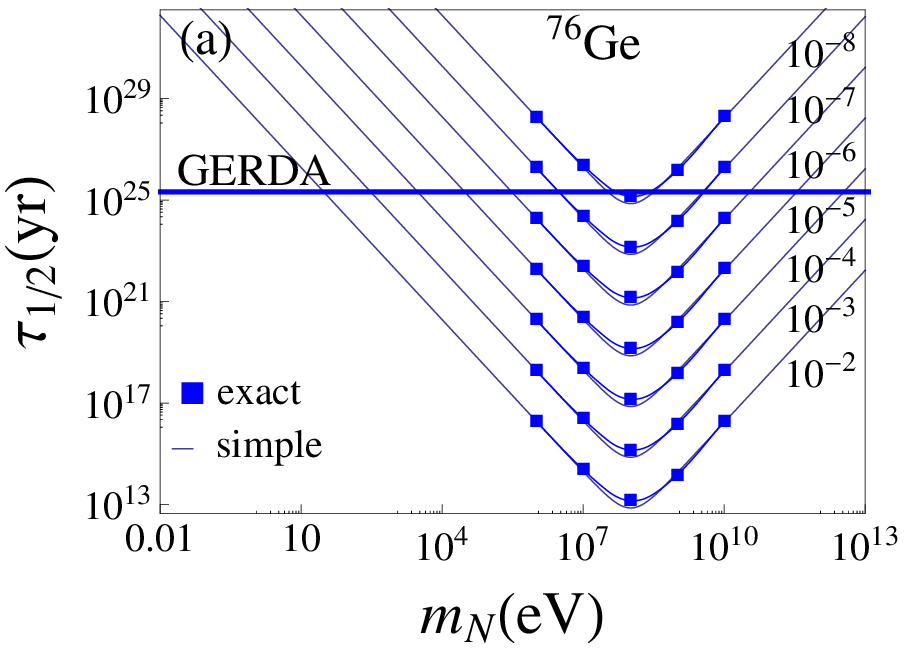} 
&\includegraphics[width=0.33\linewidth]{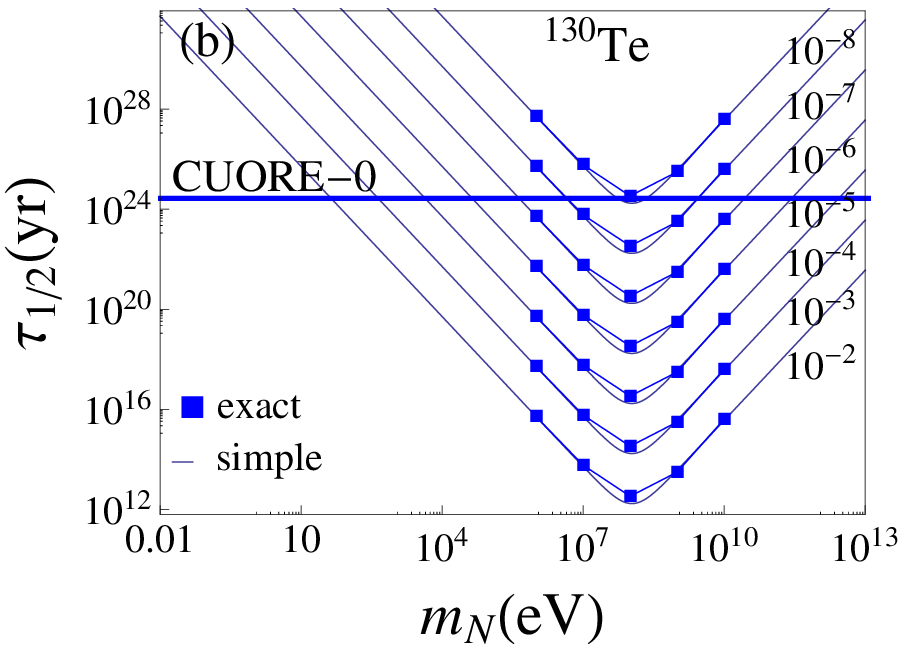} 
&\includegraphics[width=0.33\linewidth]{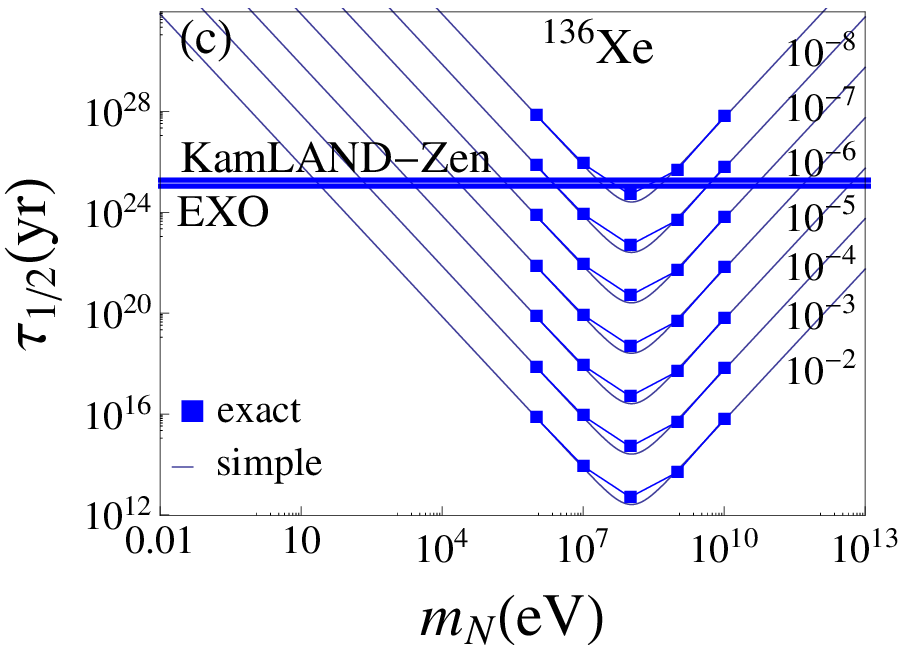} 
\end{tabular}
\caption{\label{fig2}(Color online) Expected half-life for a single neutrino of mass $m_N$ with coupling $U_{eN}^2=10^{-2}-10^{-8}$ and $g_A=1.269$ for a) $^{76}$Ge, b) $^{130}$Te, and c) $^{136}$Xe. Blue squares represent the exact calculation for $m_N=0.001$GeV, $0.01$GeV, $0.1$GeV, $1$GeV, $10$GeV. The smooth curve is obtained using the simple formula  (\ref{simple}). The experimental limits from GERDA \cite{gerda}, CUORE-0 \cite{cuore}, KamLAND-Zen \cite{kamland}, and EXO \cite{exo0nu} are also shown. The excluded zone is that below these limits.}
\end{center}
\end{figure*}

\begin{figure}[ctb!]
\begin{center}
\includegraphics[width=1.\linewidth]{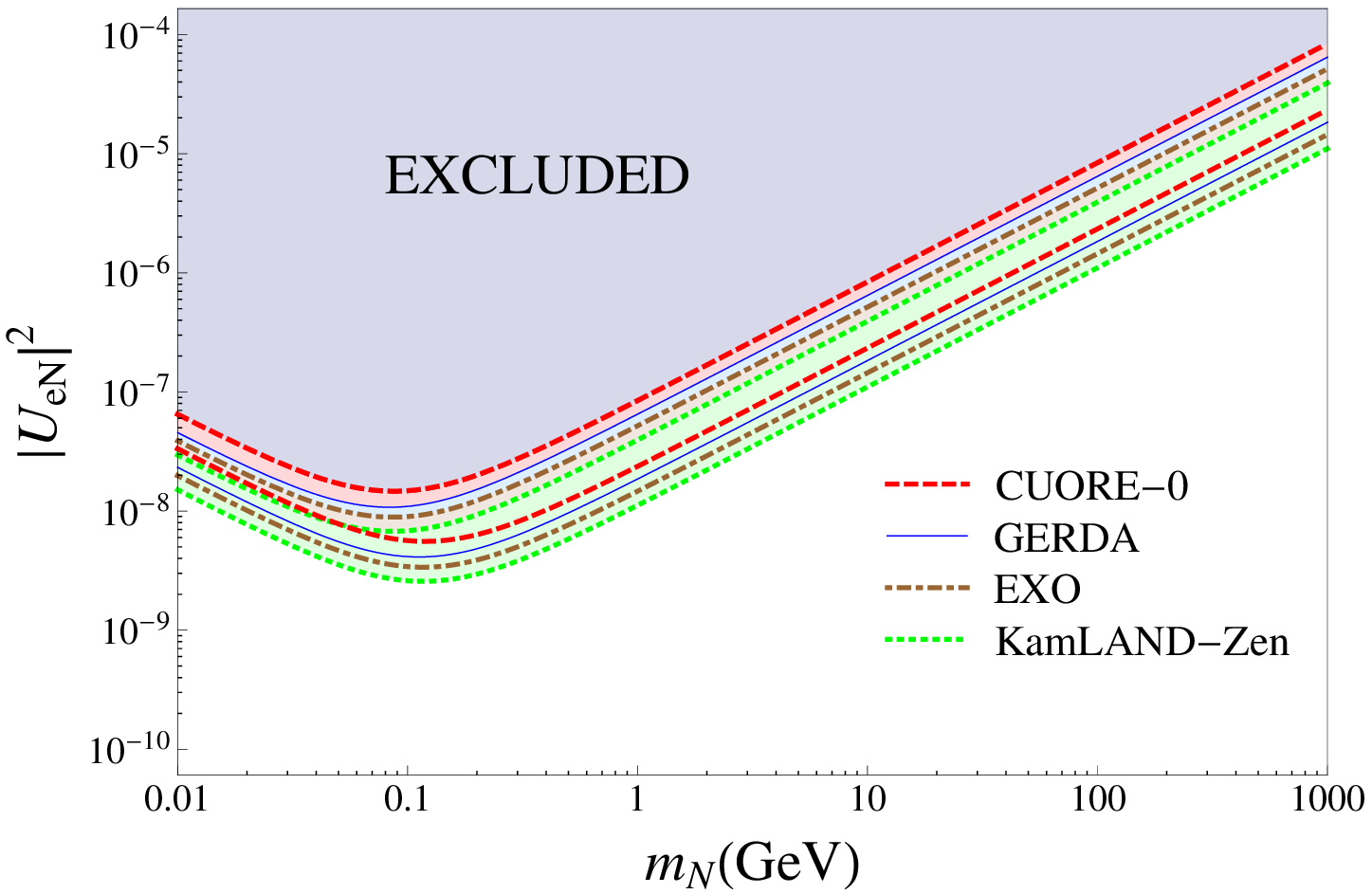} 
\caption{\label{newfig4}(Color online)Excluded values of $\left|U_{eN}\right|^2$ and $m_N$ in the $m_N$-$\left|U_{eN}\right|^2$ plane, for $g_A=1.269$. For each experiment, GERDA \cite{gerda}, CUORE-0 \cite{cuore}, KamLAND-Zen \cite{kamland}, and EXO \cite{exo0nu}, a band of values is given, corresponding to our error estimate.}
\end{center}
\end{figure}

Using Eq.~(4) and our previously calculated $G_{0\nu}$, we can calculate the expected half-life for a single neutrino of mass $m_N$ with coupling $U_{eN}$. This is shown in Fig. \ref{fig2} for several values $U_{eN}^2=10^{-2}-10^{-8}$. Note the resonant-like behavior at $m_N\sim p_F$.  For other values $U_{eN}^2$, the half-lives scale as $1/U_{eN}^4$. In the figure, the current limits from GERDA \cite{gerda} in $^{76}$Ge, CUORE-0 \cite{cuore} in $^{130}$Te, KamLAND-Zen \cite{kamland}, and EXO \cite{exo0nu} in $^{136}$Xe are also shown. The situation is summarized in the exclusion plot \cite{fae14} of Fig. \ref{newfig4}.

\begin{figure}[ctb!]
\begin{center}
\includegraphics[width=1.\linewidth]{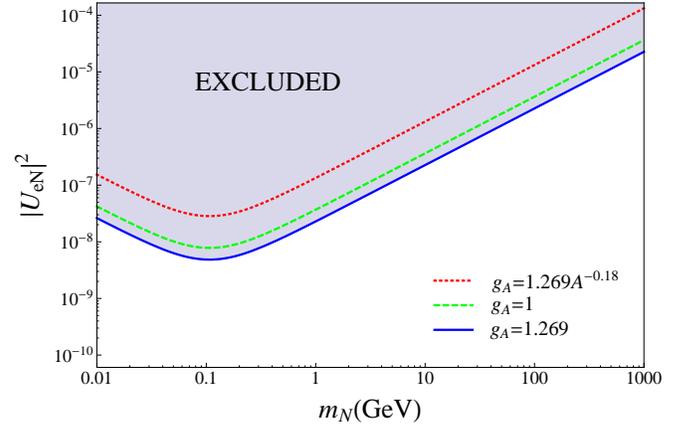} 
\caption{\label{newfig5}(Color online)Excluded region for different values of $g_A$, for $^{136}$Xe (EXO collaboration \cite{exo0nu}).}
\end{center}
\end{figure}

The calculated half-lives in Fig. \ref{fig2} are for $g_A=1.269$. For other values of $g_A$ they scale as $1/g_A^4$. The renormalization of the axial vector coupling constant, $g_A$, in nuclei is, at the present time, a major issue. Three possibilities are often considered, $g_A=1.269$ (free value), $g_A=1$ (quark value), and $g_A=1.269A^{-0.18}$ (maximal quenching). Accordingly, the excluded region varies in Fig. \ref{newfig5}. 
From this figure, one can see that if $g_A$ is renormalized in $0\nu\beta\beta$ as much as in $2\nu\beta\beta$ \cite{kot12, bar13}, and as in single-$\beta$  decay \cite{yos13,suh13,pir15}, the limits in Fig. \ref{fig2} should be multiplied by factors of $\sim 22$ in $^{76}$Ge, of  $\sim 33$ in $^{130}$Te, and of $\sim 34$ in $^{136}$Xe. 

\section{Limits on suggested sterile neutrinos}
As mentioned in the introduction several types of sterile neutrinos have been suggested. We consider here a family of neutrinos at the eV scale \cite{giu13,bar11}, a family of neutrinos at the keV scale \cite{asa05}, one at the MeV-GeV scale \cite{asa11,sha06}, and one at the TeV scale \cite{tel11}.

The total contribution to the half-life can then be rewritten, using Eq.~(\ref{simple})
\begin{widetext}
\begin{equation}
\begin{split}
\label{full}
\lbrack \tau_{1/2}^{0\nu}\rbrack^{-1}=G_{0\nu}g_A^4 &
\left\vert 
\left[ \frac{1}{m_e}\sum_{k=1}^3 U_{ek}^2 m_k+\frac{1}{m_e}\sum_i U_{ei}^2 m_i+\frac{1}{m_e}\sum_j U_{ej}^2m_j \right] M^{(0\nu)}
\right. \\
&\left.
+\left[ m_p\sum_N U_{eN}^2\frac{m_N}{\langle p^2\rangle+ m_N^2}+m_p\sum_{k_h=1}^3 U_{ek_h}^2\frac{1}{m_{k_h}}\right]
M^{(0\nu_h)} 
\right\vert ^2.
\end{split}
\end{equation}
\end{widetext}
Here we have separated the contribution of the light, $m_N\ll p_F$, neutrinos, into known $k=1,2,3$, unknown at eV scale, $i$, unknown at keV scale, $j$, and used the expression appropriate for them in terms of $M^{(0\nu)}$. We have also explicitly written the contribution of intermediate mass, $m_N\sim p_F$, neutrinos at MeV-GeV scale in the form (\ref{simple}), and finally added the contribution of heavy, $m_N\gg p_F$, neutrinos at the TeV scale, using the form appropriate for  them in terms of $M^{(0\nu_h)}$.

To set limits on sterile neutrino contribution, one should do a simultaneous analysis where all contributions are included. However, in view of the structure of Eq.~(11), it is sufficient to consider, for the purposes of this paper,  separate contributions of light, intermediate, and heavy neutrinos. This separation may, in general, not be possible, as the separate contributions may interfere with each other.

\begin{figure}[h!]
\includegraphics[width=1.00\linewidth]{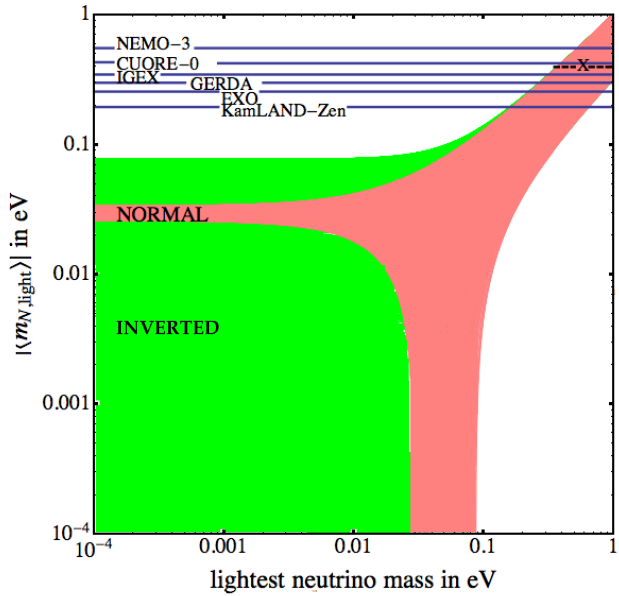}
\caption{\label{fig3new}(Color online) Current limits for $\langle m_{N,light}\rangle $ from the CUORE-0 \cite{cuore},  IGEX ~\cite{igex},  NEMO-3 ~\cite{nemo},  KamLAND-Zen~\cite{kamland}, EXO~\cite{exo0nu}, and GERDA~\cite{gerda} experiments, and IBM-2 Argonne SRC NMEs and $g_A=1.269$. The value of Ref. \cite{klapdor} is shown by $X$. The figure is in semilogarithmic scale. Red shows the normal hierarchy and green the inverted hierarchy. In this figure the scenario suggested in \cite{giu13}, relevant to LSND and reactor anomaly, is considered.}
\end{figure}

The analysis of light $(m_N\ll p_F)$ sterile neutrinos can be done as in previous publications \cite {bar13,bar15}. Introducing the quantity 
\begin{equation}
\langle m_{N,light}\rangle=\sum_{k=1}^3 U_{ek}^2m_k+\sum_{i} U_{ei}^2m_i+\sum_{j} U_{ej}^2m_j
\end{equation}
and comparing the experimental limits with the calculated half-lives using
\begin{equation}
\left[\tau_{1/2}^{0\nu} \right]^{-1}=G_{0\nu}g_A^4\left( \frac{\langle m_{N,light}\rangle}{m_e}\right)^2\left|M^{(0\nu)} \right|^2,
\end{equation}
we obtain the results in Table \ref{table3}. Considering, for example, the case suggested in  \cite{giu13} of a 4th neutrino with mass $m_4=1$eV and  $\left| U_{e4}\right|^2=0.03$, we have
\begin{equation}
\langle m_{N,light}\rangle=\sum_{k=1}^3 U_{ek}^2m_k+  U_{e4}^2e^{i\alpha_4}m_4,
\end{equation}
where the unknown phase is $0\leq \alpha_4 \leq 2\pi$. Using for the contribution of the known neutrinos the best fit values given in \cite{cap14}, we obtain the result in Fig. \ref{fig3new}. The effect of the 4th neutrino is to add to the average mass a contribution of 30meV, making the spread of the allowed values in Fig.~\ref{fig3new} larger than in the corresponding Fig.~7 of \cite{bar15} without 4th neutrino and thus improving the possibility to detect it in the next generation experiments.

The combined analysis of heavy, $m_N\gg p_F$, and intermediate, $m_N\sim p_F$, mass  neutrinos can also be done as before \cite{bar13,bar15}. The inverse half-life is given by
\begin{equation}
\left[\tau_{1/2}^{0\nu} \right]^{-1}=G_{0\nu}g_A^4\left| M^{(0\nu_h)} \right| ^2\left|\eta \right|^2
\end{equation}
with
\begin{equation}
\eta=m_p\langle m_{N,heavy}^{-1}\rangle=m_p\left[\sum_{k_h} U_{ek_h}^2\frac{1}{m_{k_h}}+\sum_{N} U_{eN}^2\frac{m_N}{\langle p^2\rangle+ m_N^2}\right].
\end{equation}
By comparing the experimental limits with the calculated half-lives, we obtain the results in Table~IV. The effect of additional intermediate mass neutrinos is to add a contribution to $\eta$.

\begin{ruledtabular}
\begin{table}[h!]
\caption{\label{table3}Left: Calculated half-lives, Eq.~(13), in IBM-2 Argonne SRC, $g_A=1.269$, and $\left< m_{N,light}\right>=1$eV. Right: Upper limit on sterile neutrino mass $\left< m_{N,light}\right>$ from current experimental limit from a compilation of Barabash \cite{barabash11}. The value reported by Klapdor-Kleingrothaus \textit{et al.} \cite{klapdor}, IGEX collaboration \cite{igex}, and the recent limits from KamLAND-Zen \cite{kamland}, EXO \cite{exo0nu}, and GERDA \cite{gerda} are also included.}
\begin{tabular}{lc|cc}
Decay  &  \ensuremath{\tau_{1/2}^{0\nu}}(\ensuremath{10^{24}}yr) &  \ensuremath{\tau_{1/2, exp}^{0\nu}}(yr) &$\left< m_{N,light}\right>$ (eV)\\
 \hline
 \T
$^{48}$Ca$\rightarrow ^{48}$Ti		&1.33 &$>5.8\times 10^{22}$ &$<4.8$\\
$^{76}$Ge$\rightarrow ^{76}$Se 	&1.95 &$>1.9\times 10^{25}$ &$<0.32$\\
							&	 	&$1.2\times 10^{25}$\footnotemark[1] &$0.40$\\
							&	 	&$>1.6\times 10^{25}$\footnotemark[2] &$<0.35$\\
							&	 	&$>2.1\times 10^{25}$\footnotemark[3] &$<0.30$\\
$^{82}$Se$\rightarrow ^{82}$Kr	 	&0.71 &$>3.6\times 10^{23}$ &$<1.4$\\
$^{96}$Zr$\rightarrow ^{96}$Mo	&0.61 &$>9.2\times 10^{21}$ &$<8.1$\\
$^{100}$Mo$\rightarrow ^{100}$Ru 	&0.36 &$>1.1\times 10^{24}$ &$<0.57$\\
$^{110}$Pd$\rightarrow ^{110}$Cd 	&1.27 & &\\
$^{116}$Cd$\rightarrow ^{116}$Sn  	&0.63 &$>1.7\times 10^{23}$ &$<1.9$\\
$^{124}$Sn$\rightarrow ^{124}$Te 	&1.09 & &\\
$^{128}$Te$\rightarrow ^{128}$Xe 	&10.19 &$>1.5\times 10^{24}$ &$<2.6$\\
$^{130}$Te$\rightarrow ^{130}$Xe 	&0.52 &$>2.8\times 10^{24}$ &$<0.43$\\
$^{134}$Xe$\rightarrow ^{124}$Ba 	&10.23 & &\\
$^{136}$Xe$\rightarrow ^{136}$Ba 	&0.74 &$>1.9\times 10^{25}$\footnotemark[4] &$<0.20$\\
								 	&	 &$>1.1\times 10^{25}$\footnotemark[5] &$<0.25$\\

$^{148}$Nd$\rightarrow ^{148}$Sm 	&1.87 & &\\
$^{150}$Nd$\rightarrow ^{150}$Sm 	&0.22 &$>1.8\times 10^{22}$ &$<3.5$\\
$^{154}$Sm$\rightarrow ^{154}$Gd 	&4.19 & &\\
$^{160}$Gd$\rightarrow ^{160}$Dy 	&0.63 & &\\
$^{198}$Pt$\rightarrow ^{198}$Hg 	&2.77 & &\\
$^{232}$Th$\rightarrow ^{232}$U 	&0.44\\
$^{238}$U$\rightarrow ^{238}$Pu 	&0.13\\

\end{tabular}
\footnotetext[1]{Ref.~\cite{klapdor}}
\footnotetext[2]{Ref.~\cite{igex}}
\footnotetext[3]{Ref.~\cite{gerda}}
\footnotetext[4]{Ref.~\cite{kamland}}
\footnotetext[5]{Ref.~\cite{exo0nu}}

\end{table}
\end{ruledtabular}

\begin{ruledtabular}
\begin{table*}[cbt!]
\caption{\label{table4}
Same as Table \ref{table3} for Eq.~(16) with $\eta=1\times10^{-7}$. The last column shows the lower limit on $\left< m_{N, heavy}\right>$.
}
\begin{tabular}{lc|ccc}
Decay  &  \ensuremath{\tau_{1/2}^{0\nu_h}}(\ensuremath{10^{24}}yr) &  \ensuremath{\tau_{1/2, exp}^{0\nu_h}}(yr) &$|\eta|(10^{-6})$ &$\left< m_{N, heavy}\right>$(GeV)\\
 \hline
 \T
$^{48}$Ca$\rightarrow ^{48}$Ti		&0.72	&$>5.8\times 10^{22}$				&<0.36		&>2.6\\
$^{76}$Ge$\rightarrow ^{76}$Se	 &1.51	&$>1.9\times 10^{25}$				&<0.028		&>33.5\\
							&	 	&$1.2\times 10^{25}$\footnotemark[1]&0.035 	&26.8\\
						 	&	 	&$>1.6\times 10^{25}$\footnotemark[2]&<0.031 	&>30.3\\
							&	 	&$>2.1\times 10^{25}$\footnotemark[3] &$<0.027$	&>34.8\\

$^{82}$Se$\rightarrow ^{82}$Kr	 	&0.55	&$>3.6\times 10^{23}$				&<0.12		&>7.82\\
$^{96}$Zr$\rightarrow ^{96}$Mo	&0.19	&$>9.2\times 10^{21}$				&<0.46		&>2.04\\
$^{100}$Mo$\rightarrow ^{100}$Ru 	&0.09	&$>1.1\times 10^{24}$				&<0.028	&>33.5\\
$^{110}$Pd$\rightarrow ^{110}$Cd 	&0.33	&									&			&\\
$^{116}$Cd$\rightarrow ^{116}$Sn  	&0.19	&$>1.7\times 10^{23}$				&<0.11		&>8.53\\
$^{124}$Sn$\rightarrow ^{124}$Te 	&0.67	&									&			&\\
$^{128}$Te$\rightarrow ^{128}$Xe 	&6.43	&$>1.5\times 10^{24}$				&<0.21		&>4.47\\
$^{130}$Te$\rightarrow ^{130}$Xe 	&0.32	&$>2.8\times 10^{24}$				&<0.034		&>27.6\\
$^{134}$Xe$\rightarrow ^{134}$Ba 	&7.73	&	&	&\\
$^{136}$Xe$\rightarrow ^{136}$Ba 	&0.50	&$>1.9\times 10^{25}$\footnotemark[4]&<0.016	&>58.6\\
								 	&		&$>1.1\times 10^{25}$\footnotemark[5]&<0.021	&>44.7\\
$^{148}$Nd$\rightarrow ^{148}$Sm 	&0.36	&									&			&\\
$^{150}$Nd$\rightarrow ^{150}$Sm 	&0.05	&$>1.8\times 10^{22}$				&<0.16		&>5.9\\
$^{154}$Sm$\rightarrow ^{154}$Gd 	&1.00	&									&			&\\
$^{160}$Gd$\rightarrow ^{160}$Dy 	&0.17	&									&			&\\
$^{198}$Pt$\rightarrow ^{198}$Hg 	&0.48	&									&			&\\
$^{232}$Th$\rightarrow ^{232}$U 	&0.11\\
$^{238}$U$\rightarrow ^{238}$Pu 	&0.03\\
\end{tabular}
\footnotetext[1]{Ref.~\cite{klapdor}}
\footnotetext[2]{Ref.~\cite{igex}}
\footnotetext[3]{Ref.~\cite{gerda}}
\footnotetext[4]{Ref.~\cite{kamland}}
\footnotetext[5]{Ref.~\cite{exo0nu}}

\end{table*}
\end{ruledtabular}

All the above limits have been set with $g_A=1.269$. As mentioned in the previous section,  $g_A$ is renormalized in models of nuclei \cite{bar13, rob13,del14}. The value of $g_A$ appears to the fourth power in the half-life. If $g_{A,eff}^{IBM-2}=1$, the quark value, all limits for $\langle m_{N, light}\rangle$ in Table III should be multiplied by a factor of $\sim 1.6$. If $g_{A,eff}^{IBM-2}=1.269A^{-0.18}$, maximal quenching, they should be multiplied by larger a factor which depends on $A$. For $^{76}$Ge, this factor is $\sim 4.7$.

\section{Conclusions}
In this article, we have calculated nuclear matrix elements for exchange of neutrinos of arbitrary mass within the framework of the microscopic interacting boson model-2 in all nuclei of interest to double beta decay. A simple analytical formula \cite{kov09} has been shown to describe the full calculation well. Using this formula and the experimental limits on half-lives, we have then set limits on possible sterile neutrino contributions in neutrinoless double beta decay. As a result, a good fraction of the parameter space in the $m_N$-$\left|U_{eN}\right|^2$ plane is excluded, especially for neutrinos in the intermediate mass range, Fig \ref{newfig4}.


\acknowledgements
This work was supported in part by U.S. Department of Energy (Grant No. DE-FG02-91ER40608), Chilean Ministry of Education (Fondo Nacional de Desarrollo Cient\'ifico y Tecnol\'ogico) (Grant No. 1150564), Academy of Finland (Suomen Akatemia) (Project 266437), and by the facilities and staff of the Yale University Faculty of Arts and Sciences High Performance Computing Center. We thank F. Vissani for useful discussions.

\end{document}